\documentclass[letterpaper,conference]{IEEEtran}
\IEEEoverridecommandlockouts
\usepackage{mathtools}
\usepackage{ifpdf}
\usepackage{amsthm}
\usepackage{nicefrac}
\usepackage{cite}
\usepackage{subfig}
\theoremstyle{definition}

\begin{document}
\title{Jamming Based on an Ephemeral Key to Obtain Everlasting Security in Wireless Environments}
\author{
    \IEEEauthorblockN{Azadeh Sheikholeslami,\IEEEmembership{ Student Member, IEEE},  Dennis Goeckel,\IEEEmembership{ Fellow, IEEE}, Hossein Pishro-Nik,\IEEEmembership{  Member, IEEE}\\
    \IEEEauthorblockA{Electrical and Computer Engineering  Department, University of Massachusetts, Amherst, MA
    \\\{sheikholesla,goeckel,pishro\}@ecs.umass.edu}\\
     \thanks{This work has been supported by the National Science Foundation
under Grants CIF-1249275 and CIF-1421957. This work appears, in part, at the 2014 Asilomar Conference on Signals, Systems, and Computers \cite{asilomar2014}.}
 }
 }

\date{}
\maketitle
\begin{abstract}
Secure communication over a wiretap channel is considered in the disadvantaged wireless 
environment, where the eavesdropper channel is (possibly much) better than the main 
channel. We present a method to exploit inherent vulnerabilities of the eavesdropper’s 
receiver to obtain everlasting secrecy. Based on an ephemeral cryptographic key 
pre-shared between the transmitter Alice and the intended recipient Bob, a random 
jamming signal is added to each symbol.  Bob can subtract the jamming signal 
before recording the signal, while the eavesdropper Eve is forced to perform
these non-commutative operations in the opposite order. Thus, information-theoretic 
secrecy can be obtained, hence achieving the goal of converting the vulnerable 
``cheap'' cryptographic secret key bits into ``valuable'' information-theoretic 
(i.e. everlasting) secure bits.  We evaluate the achievable secrecy rates for different 
settings, and show that, even when the eavesdropper has perfect access to the output 
of the transmitter (albeit through an imperfect analog-to-digital converter), the 
method can still achieve a positive secrecy rate.  Next we consider a wideband system,
where Alice and Bob perform frequency hopping in addition to adding the random 
jamming to the signal, and we show the utility of such an approach even in the 
face of substantial eavesdropper hardware capabilities.
\end{abstract}

\section{Introduction}
  The usual approach to provide secrecy is encryption of the  message. Such cryptographic approaches rely on the assumption that the eavesdropper does not have access to the key, and the computational capabilities of the eavesdropper are limited \cite{stinson2006cryptography}.  
  However, if the eavesdropper can somehow obtain the key in the future, or the cryptographic system is broken,  the secret message can be obtained from the recorded clean cipher \cite{bensonverona}, which is not acceptable in many applications requiring everlasting secrecy. 
 
The desire for everlasting security motivates considering information-theoretic security methods, where the eavesdropper is unable to extract any information about the  message from the received signal. 
 Wyner showed that,   for a discrete memoryless wiretap channel, if the eavesdropper's channel is degraded with respect to the main channel,  adding randomness to the codebook allows a positive secrecy rate to be achieved \cite{wyner1975wire}. This idea was extended to the  
  more general case of a wiretap channel with a  ``more noisy'' or ``less capable'' eavesdropper  \cite{csiszar1978broadcast}. 
Hence, in order to obtain a positive secrecy rate in a one-way communication system, having an advantage for the main channel with respect to the eavesdropper's channel is essential.
However, in wireless systems, guaranteeing such an advantage is not always possible, as an eavesdropper that is close to the transmitter or with a directional antenna can obtain a very high signal-to-noise ratio. Furthermore, the location and  channel state information of a passive eavesdropper is usually not known to the legitimate nodes, making it difficult to pick the secrecy rate to employ.
Recently, approaches based on the cooperative jamming scheme of \cite{negi2005secret} and \cite{goel2005secret}, which try to build an advantage for the legitimate nodes over the eavesdropper,  have been considered extensively in the literature \cite{lai2007cooperative,gopala2008secrecy,he2008two,dong2009cooperative,huang2011cooperative,park2013jamming}.  
 However,  these approaches require either multiple antennas, helper nodes, and/or fading and therefore are not robust across all operating environments envisioned for wireless networks.
Other approaches to obtain information-theoretic security when such an advantage does not exist are schemes based on ``public discussion'' \cite{maurer1993secret}, which utilize two-way communication channels and a public authenticated channel.  However, public discussion schemes result in low secrecy rates in scenarios of interest (as discussed in detail in \cite{jsac2013}), and the technique proposed here can be used in conjunction with public discussion approaches when two-way communication is possible. 

In this work, we exploit {\em current} hardware limitations of the eavesdropper to achieve everlasting security.  Prior work in this area includes the ``bounded storage model''  of 
Cachin and Maurer \cite{cachin1997unconditional}.
However, it is difficult to plan on memory size limitations at the eavesdropper, since not only do memories improve
rapidly as described by the well-known Moore's Law \cite{kuchibhatla2010imft}, but  they also can be stacked arbitrarily subject only to (very) large space limitations.
Our approach, first presented in \cite{allerton2012} and further developed in \cite{jsac2013}, rather than exploiting limitations of the memory in the receiver back-end, exploits the limitations of the  the analog-to-digital converter (A/D) in the receiver front-end, where the technology progresses slowly, and unlike memory, stacking cannot be done arbitrarily due to jitter considerations.   
Also,  from a long-term perspective, there is a fundamental bound on the ability to perform A/D conversion \cite{walden1999analog}, with some authors postulating that current technology is close to that limit  \cite{krone2009fundamental,krone2010fundamental}. 
Hence, we exploit the receiver analog-to-digital conversion processing effect on the received signal to obtain everlasting security.  
A rapid random power modulation instance of this approach was investigated in \cite{jsac2013} and \cite{allerton2012}, where the transmitter Alice modulates the signal by two vastly different power levels. The intended recipient Bob, since he knows the key, can adapt to the power modulation  before his A/D, while the eavesdropper Eve fails to do such and, for a restricted set of attacker modes, information-theoretic security is obtained.
However,  the power modulation scheme  is susceptible to being broken by an eavesdropper with a more sophisticated receiver than that assumed in  \cite{jsac2013}, as  discussed in \cite{jsac2013} and  shown explicitly here  in Section II.

Hence, here we consider a different method to obtain everlasting secrecy. First, recall that, at moderate-to-high signal-to-noise ratios (SNRs),
increasing the transmit power leads to very small gains in the secrecy rate, as it makes the received signal not
only stronger at Bob, but also at Eve. So, consider using excess power in a different manner. Suppose that
Alice employs her cryptographically-secure key bits to select a jamming signal  to add to the transmitted signal.
Since Bob knows the key, he can cancel the jamming signal  before his A/D; on the other hand, Eve must store the signal and try to cancel the jamming signal from the recorded signal  at the output of  her A/D after she obtains the key\footnote{Recall that the storage of an analog signal, which is equivalent to an analog delay line, is one of the greatest and longstanding challenges of analog signal processing.}.
However, the jamming signal is designed such that Eve has already lost the information she would need to recover the secret message, even if she obtains the key immediately after the transmission.
In particular, we present numerical results to investigate the number of  cryptographic key bits  needed to obtain positive secrecy rates and demonstrate the secrecy rates that can be obtained in disadvantaged environments.

Next, we consider a wideband system that additionally can employ spread-spectrum in the form of frequency hopping to further enhance everlasting secrecy.
At  first glance, one might think  that  the eavesdropper can easily thwart such an enhancement  by utilizing a wideband receiver that can process all of the frequencies that the transmitter uses.
However, because of the limitations in the aperture jitter of A/Ds,  the implementation of an A/D with both a high sampling frequency and high resolution is not feasible.
Thus, Eve faces a difficult tradeoff.
If Eve employs a high resolution A/D, she will lose any information outside her bandwidth.
On the other hand, if she employs an A/D with large bandwidth, the resolution of her A/D will be lower,
making her receiver  vulnerable to the random jamming employed by Alice, per above.

The rest of the paper is organized as follows.  Section II describes the system model and the difficult challenges faced.
The random jamming approach for secrecy, its analysis and  numerical results for the narrowband case are presented in Section III.
In Section IV,  wideband systems are considered. The method of Section IV is discussed in Section V.
Conclusions and ideas for future work are presented in Section VI.

\begin{figure}
\begin{center}
\includegraphics[width =.45\textwidth]{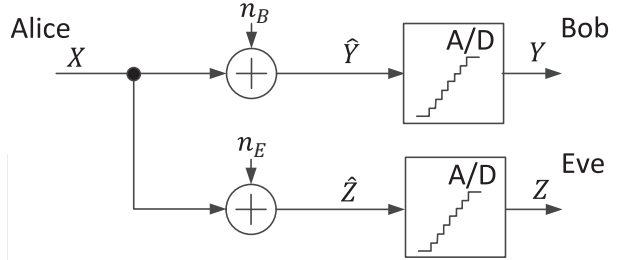}
\vspace{-5pt}
\end{center}
\caption{Wiretap channel with  receiver analog-to-digital (A/D) converters shown.
}
\label{fig:simple}
\end{figure}
\section{System Model}
\subsection{System Model}
We consider a  wiretap channel, which consists of a transmitter, Alice, a legitimate receiver, Bob, and an eavesdropper, Eve. 
 The eavesdropper is assumed to be passive, i.e. it does not attempt to actively  thwart (i.e. via jamming, signal insertion) the legitimate nodes. 
 Thus, the location and channel state information of the eavesdropper is assumed  to be unknown to the legitimate nodes. 
 We consider a  one-way communication system with an additive white Gaussian noise (AWGN) channel between Alice and each of Bob and Eve, and we include variations of the path-loss in the noise variance. 
Hence, the signal that Bob receives is:
\[\hat{Y}=X+n_B,\]
where  $X$ denotes the current code symbol, and $n_B$ is the noise of Bob's channel, $n_B\sim\mathcal{N}(0,\sigma^2_B)$. The signal that Eve receives is:
\[\hat{Z}=X+n_E,\]
where $n_E\sim\mathcal{N}(0,\sigma^2_E)$ is the noise of Eve's channel (Figure \ref{fig:simple}). 
We consider line-of-sight communication; however, the scheme works similarly on fading channels with a different calculation for the secrecy rate. 
 We assume that $X$ is taken from a standard Gaussian codebook where each entry has variance $P$, i.e. $X\sim \mathcal{N}(0,P)$.
 
  The effect of the A/D on the received signal (quantization error) is modeled by both a quantization noise, which is due to the limitation in the size of each quantization level, and missed symbols due to the quantizer's overflow. 
 The quantization noise in this case is (approximately) uniformly distributed \cite{widrow2008quantization}, so we will assume it is uniformly distributed throughout the paper.  
 For a \textit{b}-bit quantizer ($2^b$ gray levels) over the full dynamic range $[-r,r]$, two adjacent quantization levels are spaced by  $\delta={2r}/{2^b}$, and thus the quantization noise is uniformly distributed over an interval of length $\delta$. 
Throughout this paper, the quantization noise of Bob's A/D is denoted by $n_{qB}$, and the quantization noise of Eve's A/D is denoted by $n_{qE}$. 
 Quantizer overflow happens when the amplitude of the received signal is greater than the quantizer's dynamic range. We assume that Alice knows an upper bound on Eve's current A/D conversion ability (without any assumption on Eve's future A/D conversion capabilities).

\subsection{Power Modulation Approach \cite{jsac2013, allerton2012}}
\begin{figure}
\begin{center}
\includegraphics[width =.45\textwidth]{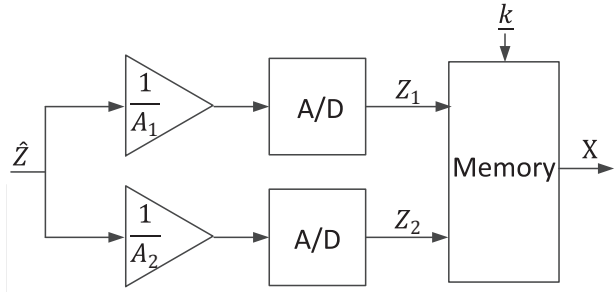}
\end{center}
\caption{Eve with a sophisticated receiver. To break the power modulation approach of \cite{jsac2013} and \cite{allerton2012}, she can record $Z_1$ and $Z_2$ and decode
them later - when she obtains the key, the encryption system is broken, or
she has access to an unlimited computational power - to obtain the secret
message.
}
\label{fig:Eve}
\end{figure}

In the scheme of \cite{jsac2013} and \cite{allerton2012}, a very short initial  key is either pre-shared between Alice and Bob, or they use a standard key agreement scheme (e.g. Diffie-Hellman \cite{diffie1976new}) to generate it. 
  This initial key will be used to generate a very long key-sequence by using a standard cryptographic method such as AES in counter mode (CTR)
  (for more details, see \cite{jsac2013,isit2013}).
 We assume that Eve
cannot recover the initial key before the key renewal and during the transmission
period. However,   we assume (pessimistically) that Eve is handed the full key (and not just the initial key) as soon as transmission is complete.  
Thus, the goal is to use the cheap (and numerous) cryptographically secure bits of the key stream to obtain ``expensive'' information-theoretic secret bits at the legitimate receiver.
 Hence, unlike  cryptographic approaches, even if the encryption system is broken later, Eve will not have enough information to recover the secret message. 

As a first step, in \cite{jsac2013,allerton2012}, we considered  a rapid power modulation instance of this approach, where the transmitted signal is modulated by two vastly different power levels $A_1$ and $A_2$ at the transmitter.   
Since Bob knows the key, he can undo the effect of the power modulation before his A/D, putting his signal in the appropriate range for analog-to-digital conversion, while Eve must compromise between larger quantization noise and more A/D overflows. Consequently, she will lose information she needs to recover the message, and information-theoretic security is obtained. 
However, a clear risk of the approach of \cite{allerton2012,jsac2013}  is a sophisticated  eavesdropper with multiple A/Ds.
Suppose that Eve has two A/Ds, and she uses them in parallel with a gain in front of each A/D such that 
each gain cancels the effect of one of the gains that Alice uses to modulate the secret message; thus, she records $Z_1$ and $Z_2$ as shown in  Figure \ref{fig:Eve}.
After completion of the transmission, if Eve obtains the key as we assume, she can use it to retain for each
channel use only the element of $\{Z_1,Z_2\}$ from the branch of her receiver properly matched to the transmission gain.  In the disadvantaged wireless scenario, Eve's recorded signal then contains more information than Bob's about the transmitted message
from Alice, and thus the desired everlasting secrecy is compromised. 
In the next section, a new approach to utilize the key bits to obtain everlasting secrecy in the case of an eavesdropper with sophisticated hardware is presented.
\begin{figure}
\begin{center}
 \includegraphics[width=.45\textwidth]{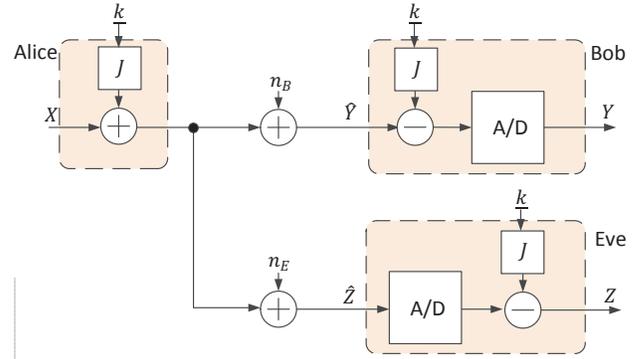}
 \end{center}
 \caption{Bob and Eve both receive the superposition of the message and the random  jamming signal.  Bob uses the key sequence to cancel the effect of the jammer on his signal before the analog-to-digital conversion, while Eve has to wait to obtain the key after completion of transmission and cancel the effect of the jammer after her A/D.}
 \label{fig:WTCwithJammer}
 \end{figure}
\section{Random Jamming for Secrecy}
In this paper, we propose adding random jamming  with large variation to the signal to obtain secrecy (Figure \ref{fig:WTCwithJammer}). 
Suppose that Alice employs her 
cryptographically-secure key bits to select a signal from a uniform discrete distribution  to add
to the transmitted signal.  
Now, since Bob knows the key, 
he can simply subtract off the jamming signal and continue normal decoding with 
an A/D converter well-matched to the span of the signal.  However, Eve 
does not have knowledge of the key and thus has difficulty
matching the span of her A/D to the received signal.  
If she does not change the span of her A/D, she will lose information due to overflows. On the other hand, if she increases the span of her A/D to contain all of the received signal, the width of each quantization level will increase and thus she will lose information due to higher  quantization noise. 
As before, we assume
that the key is handed to Eve as soon as transmission is complete,
and obviously Eve could simply subtract the jamming signal off of her {\em
recorded} samples in memory.  But, as before, a nonlinear operation
(the analog-to-digital converter) has processed the signal, hence allowing the possibility of 
information-theoretic secrecy {\em even when the secret key is 
handed to Eve immediately after the
transmission.}\footnote{We put the previous
phrase in italics so that the reader does not confuse the proposed
approach with a number of schemes in the information-theoretic 
secrecy literature that look similar, but must presume that 
the key (or secret) on which the jamming sequence is based
is kept secret from Eve forever.}
Indeed, with her poorly matched A/D,  Eve will 
not have recorded a reasonable version of the signal and we
will see  that information-theoretic security can be obtained.
In this case, one countermeasure for Eve  would be to
employ parallel receiver branches, each with a different fixed
voltage offset; however,  this is precisely a higher-resolution A/D 
over a larger span and thus is captured by the standard A/D model
and technology trend lines. 
In this paper, we will show that, through such a scheme,  ``cheap'' cryptographically-secure key 
bits can be used to greatly increase the transmission rate of the desired
``expensive'' information-theoretic secure bits.
\begin{figure}
\begin{center}
\includegraphics[width =.45\textwidth]{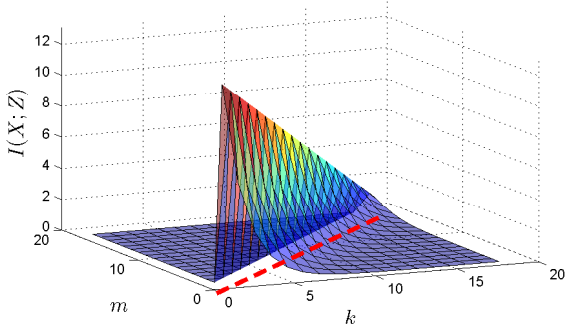}
\end{center}
\caption{$I(X;Z)$ versus $k$ (the number of key bits per jamming symbol) and $m$ (the span of Eve's A/D) when Eve has a $b_e=20$ bit A/D. Observe that  the mutual information is maximized when $m=k$ (the red dashed line).
Thus, Eve will set  the span of her A/D to $2^{k+1}l\sigma$.
}
\label{fig:span}
\end{figure}
 
 \subsection{Analysis}

 Suppose that Eve has a $b_e$ bit A/D and she sets the span of the A/D to $2l\sigma$ to cover $[-l\sigma,l\sigma]$, where $l$ is a constant that maximizes $I(X;Z)$, and $\sigma=\sqrt{P}$ is the standard deviation of the transmitted signal $X$.
Now, suppose that Alice adds a random jamming signal $J$ to $X$   based on the  pre-shared key between Alice and Bob (Figure \ref{fig:WTCwithJammer}).
 In particular, $J$ follows a discrete uniform distribution with $2^k$ levels between $-c$ and $c$, where $k$ is the number of key bits per jamming symbol,  and  $c$ (maximum amplitude of the jamming signal) is an arbitrary  constant.  
 In order to maximize the degradation of Eve's A/D, Alice should maximize $c$. Thus, given that $k$  key bits per jamming symbol is available at Alice, the relationship between $k$ and $c$ is:
$(2^k-1)\times 2l\sigma=2c$.
On the other hand, Eve, in order to maximize $I(X;Z)$,  expands the span of her A/D to $2nl\sigma $, where $n=2^m$ is an arbitrary constant that maximizes $I(X;Z)$. 
Hence, the new resolution of Eve's A/D will be
$\delta'_e=\frac{2l\sigma n}{2^{b_e}}=\frac{2l\sigma}{2^{b_e-m}}$,
and, since the jamming signal is uniformly distributed, she will miss a fraction $\frac{2^k-2^m}{2^k}$  of the  information due to her A/D overflows.
In the numerical results, we will show that the best strategy for  Eve  to employ to maximize her mutual information is to set the span of her A/D to $[-c-l\sigma,c+l\sigma]$, or equivalently $m=k$.
 Hence, in the remainder of this section, we assume the dynamic range of Eve's A/D is $2^{k+1}l\sigma$, and thus no A/D overflow happens.
In order to calculate the achievable secrecy rates, $I(X;Y)$ and $I(X;Z)$ are needed.
 We only show the calculations for the latter here, as  $I(X;Y)$ can be obtained in a similar way.
 The mutual information between $X$ and $Z$ can be written as, 
  \begin{align}
 \nonumber &I(X;Z)=h(Z)-h(Z|X)\\
 \nonumber &=\int_{-l\sigma}^{l\sigma} -f_Z(z)\log(f_Z(z))dz\\
 \label{eq:IXZ} &-\int_{-\infty}^{\infty}f_X(x)\int_{-l\sigma}^{l\sigma} -f_{Z|X=x}(z)\log(f_{Z|X=x}(z))dz dx,
    \end{align}
Hence, we need to calculate the probability density functions (pdf) of $Z$ and $Z|X=x$.
The signal at the input of Eve's receiver is 
$\hat{Z}=J+X+n_E$.
Suppose that after analog-to-digital conversion, Eve can somehow obtain the key and cancel the effect of the jamming signal. Hence, the eventual signal that  Eve obtains is:
 $Z=X+n_E+n_{qE}$.
 For simplicity of presentation, we define the random variable $Z'$ as $Z'=X+n_E$. Since $X\sim\mathcal{N}(0,P)$ and $n_e\sim\mathcal{N}(0,\sigma_e^2)$, $Z'$ follows a normal distribution with zero mean and variance $P+\sigma_E^2$.
 Hence, the probability density function of $Z$ is,
\begin{align}
\nonumber &f_Z(z)=f_{Z'}(z)*f_{n_{qE}}(z)\\
\nonumber &=\frac{1}{\delta'_E}\int_{-l\sigma}^{l\sigma}f_{Z'}(s)U_{[-\delta'_E/2,\delta'_E/2]}(z-s)ds\\
\nonumber &=\frac{1}{\delta'_E}\int_{\max(-l\sigma,z-\delta'_E/2)}^{\min(l\sigma,z+\delta'_E/2)}f_{Z'}(s)ds\\
\label{eq:fz} &=\frac{1}{\delta'_E}\left[Q\left(\frac{\max(-l\sigma,z-\delta'_E/2)}{\sqrt{P+\sigma_E^2}}\right)
-Q\left(\frac{\min(l\sigma,z+\delta'_E/2)}{\sqrt{P+\sigma_E^2}}\right)\right],
\end{align}
where $U_{[a,b]}(.)$ is the rectangle function on $[a,b]$, i.e. the value of the function is 1 on the interval  $[a,b]$ and is zero elsewhere. 

\begin{figure}
\begin{center}
\includegraphics[width =.45\textwidth]{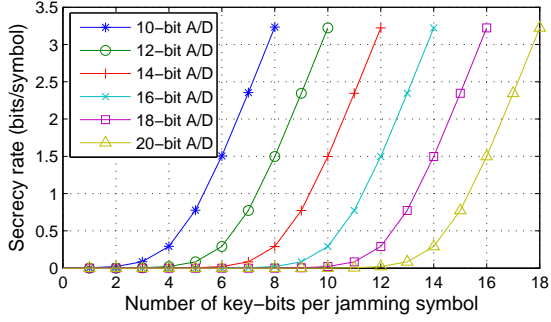}
\end{center}
\caption{Achievable secrecy rates versus the number of key bits when Bob employs a 10-bit A/D and Eve employs A/D's of various quality.  $P=1$, $l=2.5$, and the signal-to-noise ratio of each of Eve's channel and Bob's channel is 30 dB.
}
\label{fig:key}
\end{figure}

The random variable ${Z'}$ given $X=x$ has a Gaussian distribution with mean $x$ and variance $\sigma_E$. Thus, the probability density function of $Z|X=x$ is,
\begin{align}
\nonumber &f_{Z|X=x}(z)=f_{Z'|X=x}(z)*f_{n_{qE}}(z)\\\nonumber
&=\frac{1}{\delta'_E}\int_{\max(-l\sigma,z-\delta'_E/2)}^{\min(l\sigma,z+\delta'_E/2)}f_{Z'|X=x}(s)ds\\ \nonumber
&=\frac{1}{\delta'_E}\left[Q\left(\frac{\max(-l\sigma,z-\frac{\delta'_E}{2})-x}{\sigma_E}\right)\right.\\ 
 &\quad\quad\quad\quad\quad\left. -Q\left(\frac{\min(l\sigma,z+\frac{\delta'_E}{2})-x}{\sigma_E}\right)\right] \label{eq:fzx}
\end{align}
Hence, $I(X;Z)$ can be calculated by substituting (\ref{eq:fz}) and (\ref{eq:fzx}) in (\ref{eq:IXZ}). Similarly, $I(X;Y)$ can be calculated  by substituting $Z$ with $Y$, $\sigma_E$ with $\sigma_B$, and $\delta'_E$ with $\delta_B$ (where $\delta_B$ is the resolution of Bob's A/D) in (\ref{eq:IXZ}), (\ref{eq:fz}), and (\ref{eq:fzx}).
The achievable secrecy rate can be found by substituting these expressions for the mutual information into $R_s=I(X;Y)-I(X;Z)$.

In the case that the channel between Alice and Eve is noiseless (e.g. Eve picks up the transmitter), $I(X;Z)$ can be obtained from (\ref{eq:IXZ}) through the evaluation of  $h(Z)$ and $h(Z|X)$ given that the channel noise is zero. $h(Z)$ can be found by setting $\sigma_E^2=0$ in (\ref{eq:fz}), and $h(Z|X)$ can be obtained as,
\begin{align}
\nonumber 
h(Z|X)&=\int_{-\infty}^{\infty}h(Z|X=x)f_X(x)dx \\
	\nonumber&=\int_{-\infty}^{\infty}h(X+n_{qE}|X=x)f_X(x)dx\\
&=\int_{-\infty}^{\infty}h(n_{qE})f_X(x)dx
=\log(\delta'_E)
\label{eq:hzx}
\end{align}
Numerical results are presented in the next section.
\subsection{Numerical Results}
In this section,  we first show that  $I(X;Z)$ is maximized when Eve sets the span of her A/D to avoid overflow, and then we study the achievable secrecy rates of the
proposed method for various scenarios. 
In order to maximize the mutual information ($I(X;Y)$ or $I(X;Z)$), we set the quantization range by $l=2.5$ \cite{jsac2013}. 
\begin{figure}
\begin{center}
\includegraphics[width =.45\textwidth]{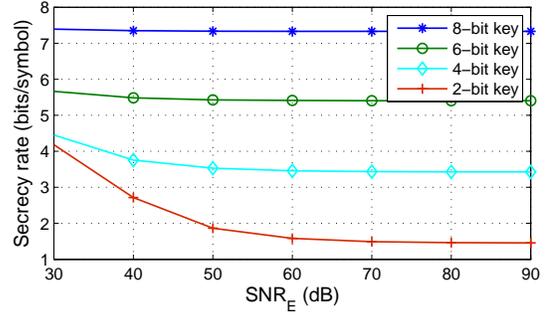}
\end{center}
\caption{Achievable secrecy rates versus the signal-to-noise ratio  of  Eve's channel ($\text{SNR}_E$) for various numbers of key bits per jamming symbol, when the SNR  of  Bob's channel is 60 dB.  $P=1$, $l=2.5$, and both Bob and Eve  use 10-bit A/Ds. Even when the quality of Eve's channel is much better than that of Bob's channel,  positive secrecy rates can be achieved.
}
\label{fig:SNRe}
\end{figure}
Since $I(X;Z)$ is an intricate function of the span of Eve's A/D ($m$) and the number of key bits employed per jamming symbol ($k$), we find the maximum of this function  numerically.
In Figure \ref{fig:span}, $I(X;Z)$ versus the span of Eve's A/D $m$ and the number of key bits $k$ when Eve employs a  $b_E=20$ bit A/D is shown.
It can be seen that the value of $I(X;Z)$ for various numbers of key bits per jamming symbol is maximized when $m=k$. 
Thus, Eve will set the dynamic range of her A/D to $2^{k+1}l\sigma$ to avoid overflow.
	
In order to see how many cryptographic key bits per symbol are needed to achieve secrecy, the curves of achievable secrecy rates versus the number of key bits per jamming symbol, for various qualities of Eve's A/D, are shown in Figure \ref{fig:key}. 
In this figure, the transmitter power $P=E[X^2]=1$, which  does not include the jamming power (Note that we will consider a total power constraint below in Figure \ref{fig:total}).
Although the quality of each of Bob's and Eve's channel is the same with a signal-to-noise ratio of 30 dB, and thus the secrecy capacity of the corresponding wiretap channel is zero,  positive secrecy rates are achieved through the proposed method. 
Further, even in the case that Eve has an A/D of much better quality than Bob's A/D (or she has stacked multiple A/Ds of the same quality as Bob's A/D until limited by clock jitter), by utilizing more key bits per jamming symbol, which are cheap cryptographic bits and can be obtained at little cost \cite{jsac2013}, positive secrecy rates (i.e. expensive information-theoretically secure bits) can be obtained.
\begin{figure}
\begin{center}
\includegraphics[width =.45\textwidth]{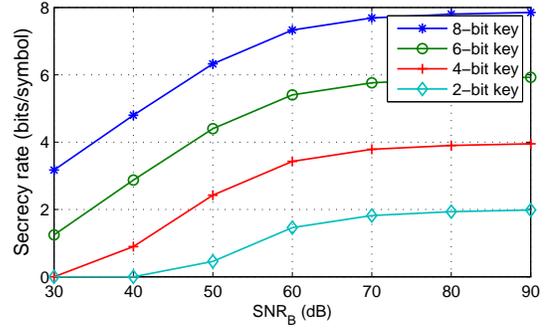}
\end{center}
\caption{Achievable secrecy rates versus signal-to-noise ratio of Bob's channel, for various numbers of key bits per jamming symbol, when Eve's channel is noiseless, i.e. Eve has perfect access to what the transmitter sends  and thus no other classical technique is effective. $P=1$, $l=2.5$, and both Bob and Eve  use 10-bit A/Ds.
}
\label{fig:noiseless}
\end{figure}
The achievable secrecy rates versus the signal-to-noise ratio  of  Eve's channel ($\text{SNR}_E$) for various numbers of key bits per jamming symbol, when the SNR  of  Bob's channel is 60 dB, are depicted in Figure \ref{fig:SNRe}. 
It can be seen that, even in  disadvantaged environments where the quality of Eve's channel is better than the quality of Bob's channel, a positive secrecy rate can be achieved.
In Figure \ref{fig:noiseless}, we look at the extreme case that Eve is able to receive
exactly what Alice transmits (e.g. the adversary is able to pick up the transmitter's radio and hook directly
to the antenna), but the channel between Alice and Bob is
noisy and hence no other classical technique\footnote{Quantum-cryptography techniques \cite{bennett1984quantum} are exempt from this.}  is effective. 
Finally, the secrecy rate versus the number of key bits per jamming symbol ($k$) for a total power constraint is shown in Figure \ref{fig:total}. 
The total power $P+P_J=1$,  Bob and Eve each have a  10-bit A/D, and  both channels have the same quality. 
When $k=0$, there is no jamming and all of the power is allocated to the signal; thus, the secrecy rate is zero. As the number of key bits (and hence the power allocated to the jamming signal) increases, the secrecy rate increases, until  eventually, as the power allocated to the portion of the signal containing the message becomes very small, it tapers at high jamming powers.

%
\begin{figure}
\begin{center}
\includegraphics[width=.45\textwidth]{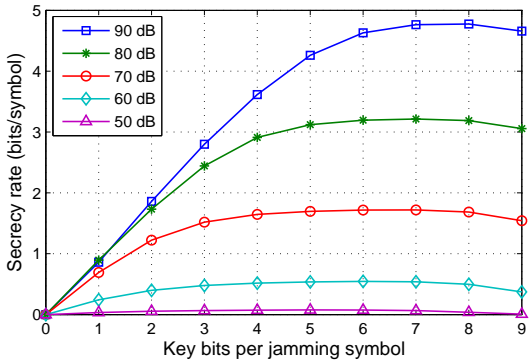}
\end{center}
\caption{Secrecy rate versus the number of key bits per jamming symbol ($k$) for various values of the total SNR, when $P+P_J=1$,  Bob and Eve  each have a  10-bit A/D, and the quality of both channels is the same.
}
\label{fig:total}
\end{figure}

 \section{Wideband Channels}
 When Alice and Bob have access to a wideband channel, they can utilize frequency hopping  based on a cryptographic key to obtain information-theoretic  secrecy.
 Frequency hopping is a spread spectrum technique that divides the entire available frequency band into sub-channels, such that at each time instance the signal is being sent over one sub-channel according to an entry in a pre-specified hopping pattern, which is known only to Alice and Bob.
 Since the hopping pattern is  not known to the eavesdropper, she will lose anything outside her receiver bandwidth, and thus this method can provide protection against eavesdropping.
 However, an eavesdropper with a wideband receiver can cover the entire frequency band.
 Motivated by the results of Section III, we will consider adding  random jamming to the signal to prevent an eavesdropper with a wide-band receiver from obtaining the message.

  Suppose that Alice is sending a narrow-band signal with bandwidth $W$ in a (large) frequency band of span $F$ by employing a narrowband slot centered at a randomly chosen center frequency $f_c$.
	Since Bob knows the key, he can tune his receiver to the correct frequency slot, while Eve is forced to either expand the bandwidth of her receiver or risk missing symbols.

\begin{figure}
\centering
\subfloat[Frequency hopping]{\includegraphics[width=.5\textwidth]{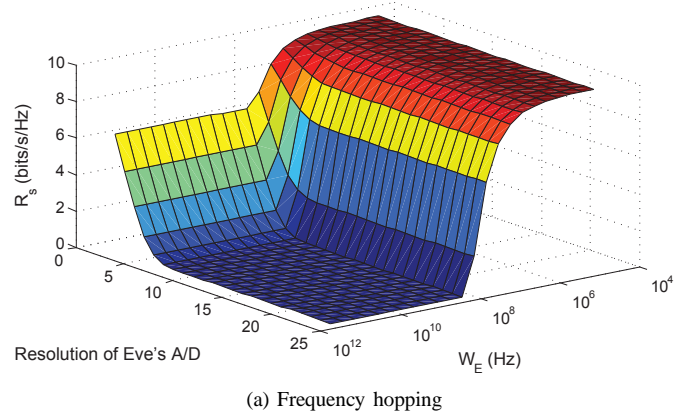}
}\\
\subfloat[Frequency hopping and random jamming]{\includegraphics[width=.5\textwidth]{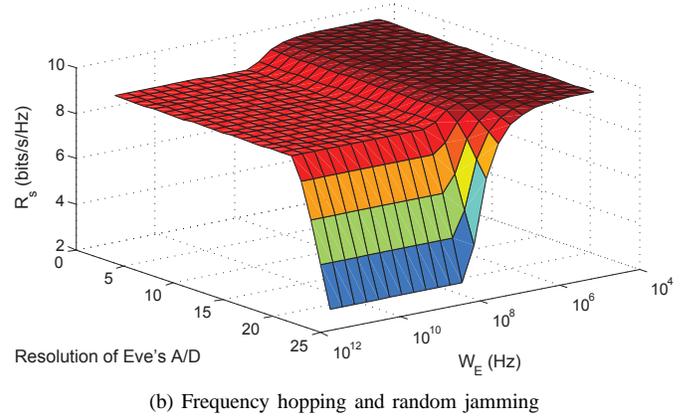}
}
\caption{Secrecy rate versus bandwidth and resolution of Eve's receiver: (a) frequency hopping, (b) frequency hopping with a 20-bit random jamming signal. Both Bob and Eve have  channels of SNR equal to 30 dB. Bob has a 20-bit A/D, signal bandwidth $W=100$ kHz,  and transmitter span $F=100$ MHz. Without random jamming, the worst case secrecy rate is zero, while with random jamming, a worst case secrecy rate of 2.736 bits/s/Hz is achievable.}
\label{fig:rs_vs_b_r}
\end{figure}

  Let $Z_B$ be the total additive noise, including both channel and quantization noise, at Bob's receiver and $Z_E$ be the total additive noise at Eve's receiver.
	In the case of band-limited channels, since each of $Z_B$ and $Z_E$ is a superposition of a Gaussian noise (channel) and a uniform noise (quantization),  the capacity of Bob's channel and Eve's channel cannot be found directly.
	Fortunately, upper and lower bounds on the capacity of a band-limited channel with independent additive noise are available \cite{shannon1949communication,iha1978cap}.
	For a channel with bandwidth $W$, for a  signal power of $P$ and total additive noise $Z$ with power $N$, the capacity $C$ lies between  bounds,
	\begin{equation}\label{eq:shannon}
W\log\left(\frac{P+N}{N}\right)\leq C\leq W\log\left(\frac{P+N}{\mathcal{E}_Z}\right),
\end{equation}
where $\mathcal{E}_Z$ is the \textit{entropy power} of $Z$. 
Suppose $Z$ is a superposition of Gaussian noise $G$ and uniform noise $U$. Since $G$ and $U$ are zero-mean  independent random variables, 
\begin{equation}\label{eq:var}
  N=E[Z^2]=Var(G)+Var(U)=\sigma^2+\frac{\delta^2}{12}.
  \end{equation}
The entropy power for a random variable $Z$ is defined as,
\[\mathcal{E}_Z=\frac{1}{2\pi e}2^{2h(Z)},\]
where $h(Z)$ is the differential entropy of $Z$.
Here, the entropy power of $Z$ cannot be easily calculated.
Hence, we use the convolution  inequality of entropy powers to find an upper bound on $\mathcal{E}_Z$.
The convolution inequality of entropy powers states that the entropy power of the sum of two independent  random variables is greater than or equal to the sum of the  entropy powers of the summands \cite{shannon1949communication,bla1965con}. Thus,
\begin{align}\nonumber \mathcal{E}_Z &\geq \mathcal{E}_G+\mathcal{E}_U\\
\nonumber &=Var(G)+\frac{6}{\pi e}Var(U)\\
&=\sigma^2+\frac{\delta^2}{2\pi e}.\label{eq:enpower}
\end{align}
From  (\ref{eq:shannon}) and (\ref{eq:var}), the capacity of Bob's channel can be lower bounded as:
\begin{align}\nonumber
C_B&\geq W\log\left(\frac{P+N_B}{N_B}\right)\\
&= W\log\left(\frac{P+\sigma^2_{B}+\frac{\delta_B^2}{12}}{\sigma^2_{B}+\frac{\delta_B^2}{12}}\right).\label{eq:cb2}
\end{align}
Now consider Eve's receiver with bandwidth $W_E$. If $W_E<F$, since Eve does not know the hopping pattern,  she will lose anything that is sent outside of the bandwidth that she is currently monitoring, i.e. she can only obtain a fraction $\frac{W_E}{F}$ of the message. Using this fact and from (\ref{eq:shannon}), (\ref{eq:var}) and (\ref{eq:enpower}), an upper bound for the capacity of Eve's channel is:
\begin{align}\nonumber
C_E&\leq \frac{W_E}{F}W\log\left(\frac{P+N_E}{\mathcal{E}_{ZE}}\right)\\
&\leq \frac{W_E}{F}W\log\left(\frac{P+\sigma^2_{E}+\frac{\delta_E^{2}}{12}}{\sigma^2_{E}+\frac{\delta_E^{2}}{2\pi e}}\right).\label{eq:ce2}
\end{align}
Thus, from (\ref{eq:cb2}) and (\ref{eq:ce2}), a lower bound on the secrecy capacity is:
\begin{align}
\nonumber C_s&=C_B-C_E\\
\label{eq:cs1} &\geq W\log\left(\frac{P+\sigma^2_{B}+\frac{\delta_B^2}{12}}{\sigma^2_{B}+\frac{\delta_B^2}{12}}\right)-\frac{W_EW}{F}\log\left(\frac{P+\sigma^2_{E}+\frac{\delta_E^{2}}{12}}{\sigma^2_{E}+\frac{\delta_E^{2}}{2\pi e}}\right).
\end{align}
Hence, any secrecy rate,
\begin{align}\nonumber
R_s= &\left[W\log\left(\frac{P+\sigma^2_{B}+\frac{\delta_B^2}{12}}{\sigma^2_{B}+\frac{\delta_B^2}{12}}\right)\right. \\ \label{eq:rs1}
&\qquad\left. -\frac{W_EW}{F}\log\left(\frac{P+\sigma^2_{E}+\frac{\delta_E^{2}}{12}}{\sigma^2_{E}+\frac{\delta_E^{2}}{2\pi e}}\right)\right]^+,
\end{align}
is achievable.
The secrecy rate versus bandwidth and resolution of Eve's receiver is shown in Figure \ref{fig:rs_vs_b_r}. In this example, both Bob and Eve have  channels with SNR equal to 30 dB. Bob has a 20-bit A/D, $W=100$ kHz,  and transmitter span $F=100 $ MHz. In Figure \ref{fig:rs_vs_b_r}(a) frequency hopping with a 10-bit key is employed, and it can be seen that by using a wideband receiver and a high resolution A/D at Eve, the secrecy rate is zero.
In Figure \ref{fig:rs_vs_b_r}(b), in addition to the frequency hopping, a random jamming signal using a 20-bit key is added to the signal, which helps the legitimate nodes to obtain a positive secrecy rate (for this setting, the worst case rate of  2.736 bits/s/Hz is available), even when Eve uses a wideband receiver and a high resolution A/D. 
However, in this case the number of jamming bits needed to gain a non-zero secrecy rate is large.
Furthermore, if Eve uses a wideband receiver with a very high resolution A/D, the secrecy capacity  with a 20-bit A/D at Bob will be zero.
In the next section, we will see that speed-resolution limitations of A/Ds  help us to obtain non-zero secrecy rates with much fewer numbers of key bits per jamming symbol, and in all feasible scenarios.
\section{Discussion}
\begin{figure}
\centering
{\includegraphics[width=.5\textwidth]{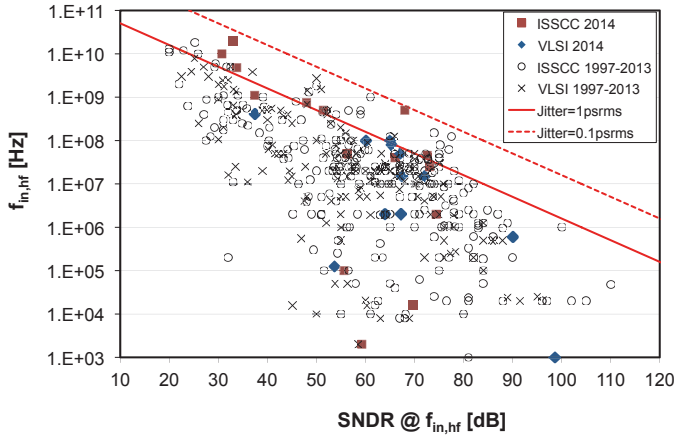}}
\caption{Sampling frequency versus SNDR of published works from 1997 to 2014. The red lines are  jitter envelopes. The solid line corresponds to 1 ps aperture jitter and the dashed line corresponds to 0.1 ps aperture jitter (courtesy of \cite{mur2014}).}
\label{fig:aperture}
\end{figure}

A limiting factor in the performance of A/Ds is aperture jitter, which is the  uncertainty in the sampling time of the A/D. 
In order to better understand the relationship between the aperture jitter of an A/D  and its impact on an A/D's performance, let us consider the signal-to-noise-and-distortion ratio (SNDR) of an A/D.
The SNDR of an A/D is the ratio of the root mean square (rms) of the  amplitude of the input signal to the rms sum of all other spectral components.
The signal-to-noise-and-distortion ratio due to the aperture jitter $t_j$ when the input signal is sampled with frequency $f_{in}$ is \cite{kester2005data}:
\begin{equation}\label{eq:SNDRj}
\text{SNDR}_j=20 \log_{10}\left(\frac{1}{2\pi f_{in} t_j}\right).
\end{equation}
Hence, when other non-idealities (quantization noise, thermal noise, etc.) of the A/D are not considered,  
(\ref{eq:SNDRj}) describes the performance limit of an A/D  due to the aperture jitter.
The sampling rate versus SNDR for trends in the current state-of-the-art of A/Ds is shown in  Figure \ref{fig:aperture}.
In this figure, the performance envelope of (\ref{eq:SNDRj}) is shown by a solid red line for $t_j=1 $ ps, and a dashed red line for $t_j=0.1$ ps.
It can be seen that the best aperture jitter achieved by the current state-of-the art of A/Ds is 0.1 ps.

The relationship between SNDR and effective number of bits (ENOB) can be found by  using the relationship between SNDR and number of bits for an ideal A/D,
\begin{equation}\label{eq:ENOB}
\text{SNDR}=6.02\:\text{ENOB}+1.76\: \text{dB}
\end{equation} 
Thus, in Figure \ref{fig:aperture}, with a change in the scaling of the horizontal axis, SNDR is equivalent to ENOB.

From the numerical results  at the end of the previous section, it seems that it is always possible for  Eve to use  a wideband high-resolution A/D to compromise secrecy, in the same way that a larger than envisioned memory at Eve would compromise secrecy in the bounded memory model of \cite{cachin1997unconditional,cachin1998oblivious,aumann1999information,dziembowski2002tight,aumann2002everlasting,dziembowski2008bare,dowsley2014oblivious}.
However, from the discussion above, aperture jitter is a critical limitation in the performance of A/Ds  that prevents Eve from increasing the bandwidth and resolution of her A/D arbitrarily.
In fact, aperture jitter restricts the product of the sampling rate and the resolution of an A/D.

\begin{figure}
\begin{center}
\includegraphics[width=.5\textwidth]{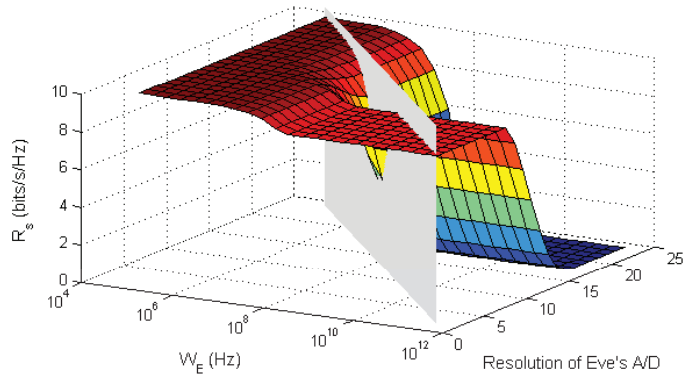}
\end{center}\vspace{0pt}
\caption{Secrecy rate versus bandwidth and resolution of Eve's receiver, when a 10-bit random jamming signal is added to the transmitted signal. Both Bob and Eve have channels with SNR of 30 dB. Bob has a 20-bit A/D, $W=100$ kHz,  and transmitter bandwidth $F=100 $ MHz.
The gray plane  is the current technology jitter envelope (0.1 ps). The bandwidth-resolutions on the right side of the gray plane are not feasible by the current technology, and thus a worst case secrecy rate of 4.25 bits/s/Hz is available.
}\vspace{0pt}
\label{fig:surf}
\end{figure}

In Figure \ref{fig:surf}, the secrecy rate versus bandwidth and resolution of Eve's A/D, when a 10-bit random jamming signal is added to the transmitted signal  is shown.
In this figure, both Bob and Eve have  channels with an SNR of 30 dB. Bob has a 20-bit A/D, a signal bandwidth $W=100$ kHz,  and a transmitter span of  $F=100$ MHz.
The gray plane in this figure is the current technology jitter envelope.
Assuming that the eavesdropper has access to the best A/D with current technology, the bandwidth-resolutions she can utilize are restricted to  this jitter envelope, and anything beyond this envelope is not feasible. 
Hence, the minimum secrecy rate in this case is 4.25 bit/s/Hz.
This shows that, in practice,  much fewer  key bits per jamming symbol are needed compared to Figure \ref{fig:rs_vs_b_r}(b), and, for current technology (0.1 ps jitter),  the eavesdropper cannot compromise secrecy  by employing a better A/D.
Clearly, in this case either the aperture jitter of the eavesdropper's A/D should be known to legitimate nodes, or they should assume that the eavesdropper has access to the best current technology.

\subsection{Aperture jitter evolution and its ultimate limit}
The best aperture jitter for A/Ds  has changed from 100 ps in the 1980s to 0.1 ps in the current state of the art.
This improvement of the jitter might seem unfavorable to  our proposed method, as we rely on the non-ideality of the eavesdropper's A/D. 
Thus, if the jitter improves with the same slope,  unbounded over  time, it can destroy the ability to transmit messages with  the proposed method at some point in the future.
However, the trend of A/D jitter shows that the current state of the art was achieved in 2005, and has not changed significantly since that time.
This, along with the fact that the technology scaling has  changed dramatically since 2005, suggests that the performance of  A/D aperture jitter is already in a state of saturation \cite{jons2012d,mur2014}.
Nevertheless, it is possible that the aperture jitter improves over  time.
Fortunately, there exists an ultimate limit on the ability to store an accurate reconstruction of an analog signal due to the Heisenberg uncertainty principle  \cite{walden1999analog,krone2009fundamental,krone2010fundamental}.

\section{Conclusion and Future Work}
In this paper, we have introduced a method to convert  ephemeral ``cheap'' cryptographic key bits to ``expensive'' information-theoretically secure bits to achieve everlasting security in  wireless environments where the intended recipient Bob is at a disadvantage relative to the eavesdropper  Eve. 
A random jamming signal chosen from a discrete uniform random ensemble based on a key pre-shared between the transmitter Alice and Bob is added to each transmitted symbol.
 The intended receiver uses the key sequence to subtract the jamming signal, while the eavesdropper Eve, in order to prevent  overflows of her analog-to-digital (A/D) converter, needs to enlarge her A/D span and thus degrade the resolution of her A/D, thus resulting in
information loss even if Eve is handed the key at the conclusion of transmission and is able to modify her recorded signal to attempt to remove the jamming effect.  The numerical results show that this method can  provide secrecy even in the case that the eavesdropper has perfect access to the output of the transmitter's radio and an A/D of much better quality than that of the intended receiver. 

We have also considered the case when  the legitimate nodes and the eavesdropper have access to wideband channels. 
When the channel bandwidth is larger than the signal bandwidth, the legitimate nodes can use their cryptographic key bits to try to hide the location of the signal from Eve by employing a frequency hopping technique.  
With this extra degree of freedom, since the product of the bandwidth and the resolution of an A/D is limited by its  aperture jitter, Eve will be forced to make an additional tradeoff between A/D resolution and sampling frequency.
Hence, the strategy of system nodes is to use frequency hopping in conjunction with the additive random jamming method.
Eve must choose to either use a high resolution A/D with small bandwidth and thus lose anything outside her bandwidth, or use 
a wideband A/D with low resolution and thus be susceptible to the random jamming signal. 
Technology trend lines and fundamental limits for A/Ds indicate this will pose a significant challenge to Eve.

\bibliographystyle{ieeetr}
\bibliography{mycite}
\end{document}